\documentclass[prd,showpacs,showkeys,superscriptaddress,amsmath,amssymb,10pt]{revtex4}
\begin{document}

\title{Spacetime as a deformable solid}
\author{M. O. Tahim}
\email{mktahim@fisica.ufc.br} \affiliation{Departamento de F\'{\i}sica - Universidade Federal do
Cear\'{a} \\ C.P. 6030, 60455-760 Fortaleza-Cear\'{a}-Brazil}

\affiliation{Departamento de Ci\^{e}ncias da Natureza, Faculdade de Ci\^{e}ncias, Educa\c{c}\~{a}o
e Letras do Sert\~{a}o Central (FECLESC), Universidade Estadual do Cear\'{a}, 63900-000
Quixad\'{a}-Cear\'{a}-Brazil}

\author{R. R. Landim}
\email{renan@fisica.ufc.br} \affiliation{Departamento de F\'{\i}sica - Universidade Federal do
Cear\'{a} \\ C.P. 6030, 60455-760 Fortaleza-Cear\'{a}-Brazil}

\author{C. A. S. Almeida}
\email{carlos@fisica.ufc.br} \affiliation{Departamento de F\'{\i}sica - Universidade Federal do
Cear\'{a} \\ C.P. 6030, 60455-760 Fortaleza-Cear\'{a}-Brazil}

\begin{abstract}
In this letter we discuss the possibility of treating the spacetime by itself as a kind of
deformable body for which we can define an fundamental lattice, just like atoms in crystal
lattices. We show three signs pointing in that direction. We simulate the spacetime manifold by a
very specific congruence of curves and use the Landau-Raychadhuri equation to study the behavior
of such a congruence. The lattice appears because we are forced to associate to each curve of the
congruence a sort of fundamental "particle". The world-lines of these particles should be
identified with the congruence fulfilling the spacetime manifold. The conclusion is that when
describing the deformations of the spacetime the Einstein equations emerge and the spacetime
metric should be treated as a \textit{secondary (not fundamental) object of the theory}.
\end{abstract}



\maketitle

\section{Introduction}

The idea pointing out some resemblances between the spacetime and
the dynamics of deformable bodies is not new. There are authors that
have discussed these observations some decades ago. The idea
generally discussed is related with the real meaning of the
spacetime: can it be quantized in the sense we generally work in
these modern days (the canonical approach and the others methods)?
If the spacetime is some sort of fluid, how can we identify its
fundamental constituents (a real crystal, from the macroscopic
viewpoint, looks like a continuum, but from the microscopic
viewpoint, is made of small parts, i.e., atoms and molecules)? If
these ideas are reliable then perhaps it does not make sense to
quantize the spacetime using the standard procedures. In this way
the spacetime as we observe and describe would be a secondary
entity, originating from an collectivity of more fundamental objects.
The first realization of these ideas has appeared with the
Sakharov's work \cite{sakharov} related with what is known as
Emergent Gravity. Following this line, the Einstein's gravitational
theory would appear after a direct quantization of matter. The
dynamics of the gravitational field is generated as a secondary
effect associated to radiative corrections at one loop. Several
lines of research have started through the years derived from
Sakharov's idea (an example is the so called Stochastic Gravity
\cite{gravity_estocastica}). On the other hand, in the last years a
Ted Jacobson's paper have shown how to understand the Einstein's
equations from a thermodynamical viewpoint \cite{jacobson}: the
Einstein's equations are equations of state for the spacetime. That
conclusion strongly suggests that the spacetime may really be
compared to a special kind of deformable body. It is just in these
lines that we propose in this work to analyze a total of three signs
pointing in that direction, i.e, that the spacetime is a kind of
deformable body. The first sign is related to the deformations of
the spacetime. The second one shows an "elastic" origin for the
Einstein-Hilbert action, and the third one shows a relation between
the Newton's law of gravitation and the Hooke's law of elasticity.
In order to discuss these signs, we adopt the spacetime as modeled
by a congruence of very specific curves: at all points of the
spacetime manifold, (it is defined by such a curve). The characteristics
of these curves can be studied by the use of the Landau-Raychaudhuri
equation \cite{lan,ray,wald} revealing if the spacetime, as modeled in this
way, is or is not a curved manifold. It is important to note that is
very common in physics to study a continuum system by discretizing
it. In canonically quantizing a field theory, for example, we treat
the field as a collection of interacting harmonic oscillators and
postulate some commutation rules in order to construct the physical
spectrum. How to do this for spacetime itself? The usual procedure
is based on the metric tensor field $g_{\mu\nu}$. We again just try
the same idea of discretization of this field. The result everybody
know: Gravity is nonrenormalizable. Nevertheless, note that
$g_{\mu\nu}$ does not represent spacetime by itself. It is a field
that depends on the \textit{coordinates} of the spacetime. An example may
help. Consider a solid with a well defined lattice. We know that
there are some excitations of this solid called "phonons". These
objects can be studied by a scalar field theory in the continuum
limit. In this case, what is the "spacetime"? We can choose a single
atom of the lattice as reference frame and in this way we get time
and distance information. The scalar field as a collective
excitation of the lattice will depend on distance and time related
to that reference frame. Then we claim that spacetime is the own
lattice and from the microscopic viewpoint it is discrete because
the lattice is made of "atoms". The conclusion is that modeling the
spacetime manifold by a congruence of curves requires us to
associate to each such a curve a fundamental "particle". The
world-lines of such particles will compose the congruence fulfilling
the spacetime manifold. Let us continue with the example of the
solid and its lattice. Because the lattice is made of several atoms
we know how to compute distances, i. e., we have a metric defined at
every point in the continuum limit. Now which field is more
fundamental, the metric one or the scalar field describing the
lattice excitations? We claim that the more fundamental field is the
scalar field because it is related directly with the atoms composing
the lattice. These atoms are true reference frames in the sense that
they are just matter reference frames: the physical notion of
distance comes after the notion of matter, then the metric is a
secondary object. In what follows we will consider the spacetime as
a kind of solid for which we can define such a lattice. In field
theory we already do this. We regard gauge fields, the metric tensor
field and spinor fields as fundamental fields defined at every point
of spacetime. The approach we will follow here is to consider any
field except the metric tensor field as fundamental and this field
will give us the notion of "spacetime" in the sense discussed above.
The consequence of this idea is that if we want to study the
deformations of such a spacetime we arrive at the Einstein's
equations as emergent equations. The organization of this work is
the following: in the first section we make a review of the
Landau-Raychaudhuri equation and discuss congruences of curves. The
second section deals with the signs of the spacetime as an elastic
body. Finally, we discuss the meaning of these signs and
perspectives.

\section{The Landau-Raychaudhuri Equation}

The Landau-Raychaudhuri equation is an equation that
describes the behavior of a congruence of timelike or spacelike
curves. It is commonly used to study and establish singularities of
the spacetime. Consider a congruence of curves fulfilling the
spacetime manifold. If there are expansion, distortion and relative
rotation between the curves of the congruence then the spacetime
manifold described by that congruence is curved. For the purposes of
this work we will be using the Landau-Raychaudhuri equation for
timelike or spacelike curves, depending on what sign we will be
discussing. So, we have respectively
\begin{equation}
\frac{d\Theta }{d\tau }=-\frac{1}{3}\Theta
^{2}-\left( \sigma ^{ab}\right)
^{2}+\left( \varpi ^{ab}\right) ^{2}-R_{cd}\xi
^{c}\xi ^{d}
\end{equation}
and
\begin{equation}\label{eq1}
\frac{d\Theta }{d\lambda
}=-\frac{1}{2}\Theta ^{2}-\left( \widehat{\sigma }%
^{ab}\right)
^{2}+\left( \widehat{\varpi }^{ab}\right) ^{2}-R_{cd}\kappa ^{c}\kappa
^{d},
\end{equation}
where $\tau $, $\lambda $ are the parameters used to describe the curves
of the congruences. $\xi ^{c}$ and $\kappa ^{c}$  are the tangent vectors
to the curves (generators of the congruence) and they are, in that order,
a timelike and a lightlike vector. $\Theta $ is a scalar and describes
expansion of volume, $\sigma ^{ab}$/$\widehat{\sigma }^{ab}$ measures the distortion of volume and $\varpi ^{ab}$/$\widehat{\varpi }^{ab}$ measures  the
rotation of the curves. In the case of this work, we will be interested in small
distortions of volume in such a way that the quadratic terms in the Landau-Raychaudhuri
equation may be disregarded (they are like second order corrections). In these
conditions, the Landau-Raychaudhuri equation can be easily integrated giving the scalar of expansion as a function of the Ricci tensor:
\begin{equation}\label{eq2}
\Theta =-\tau R_{cd}\xi
^{c}\xi ^{d}\equiv -\lambda R_{cd}\kappa
^{c}\kappa
^{d}.
\end{equation}
The objects that appear in the Landau-Raychaudhuri equation
can be obtained from a kinematical decomposition of general tensors.
A general tensor can be decomposed into a symmetric plus an antisymmetric
part:
\begin{equation}\label{eq3}
B_{ab}=B_{\left(
ab\right) }+B_{\left[ ab\right] }.
\end{equation}
The antisymmetric part is associated with the measure of rotation of
the congruence. The symmetric part can yet be decomposed into a
trace (the scalar of expansion) and a symmetric traceless piece which
is associated with the measure of the distortion of volume. The full
symmetric part will be identified with the tensor of deformation of
spacetime in analogy with the case of mechanics of deformable
bodies. In the sections that follows the spacetime will be simulated
by a congruence of timelike or lightlike curves, depending on the
sign we will be discussing.

\section{First sign: The deformations of the spacetime}

Consider a small region of the spacetime containing the point $P$.
This region defines the volume element $dV$. The question here is
how to study the deformations of this volume caused by some
external agent? One way to do this is by the use of the Lie
derivative. In fact, if we get a volume $V$ and we propagate it
using a congruence of integral curves we will obtain the modified
volume $V^{\prime }$. The difference between these two configurations
gives us a way to measure the total deformation. With this in mind we
postulate the following action for points of the spacetime:
\begin{equation}\label{eq4}
S=k\int dV.
\end{equation}
The volume described above is just the Riemannian volume form,
invariant under general coordinate transformations. The constant $k$
has the necessary dimension in order to give the correct dimension
for the action, which is "energy$\times$time". This action is quite
different from the usual actions in field theory because there is no
"a priori" lagrangian density. The usual actions carry information
about energy due to some fields distributed along some regions of
spacetime and it seems that it does not happen with our proposed
action. This is not a problem if we remember the idea discussed in
the introduction above: the volume actually comprises part of the
lattice associated with some field. This means that the volume of
this spacetime carries energy associated with the lattice. Now, we
minimize that action in the following manner: we take its Lie
variation and requires that it should be stationary, just like the
usual procedure. Then, the equation of motion for the points of that
specified region is
\begin{equation}\label{eq5}
\delta _{Lie}S=k\int
\Theta dV=0,
\end{equation}
where $\Theta $ is the scalar of expansion associated to the volume
$dV$. Using the Landau-Raychaudhuri equation with the conditions
cited in the first section, i. e., in a situation where $\sigma
^{ab}=\varpi ^{ab}=0$, then we have $\Theta =-\lambda
R_{ab}\kappa^{a}\kappa ^{b}$. Substituting this result back in the
equation of motion above we obtain
\begin{equation}\label{eq6}
-k\int \lambda
R_{ab}\kappa ^{a}\kappa ^{b}dV=0.
\end{equation}
We can establish the equality above for all null vector $\kappa ^{a}$
(we assume here the congruence is null-like which means the
fundamental particles are massless), i.e.,
\begin{equation}\label{eq7}
R_{ab}=f(g)g_{ab},
\end{equation}
where $f(g)$ is a function that just depends on the metric of the
spacetime. This function can be easily found by requiring the
disappearance of the covariant divergence of the last equation (it
is like a type of "conservation of deformation") which leads to the
result
\begin{equation}\label{eq8}
R_{ab}-\frac{1}{2}Rg_{ab}\pm \Lambda g_{ab}=0,
\end{equation}
which is just the side corresponding to the geometry of the spacetime
in the Einstein's equations. Conclusion: when we deform the volume
of the spacetime, \textit{the Einstein's equations give us a way to
understand such deformations}.

\section{Second sign: The elastic origin of the Einstein-Hilbert action}

Consider now the following functional which we will identify as an action:
\begin{equation}\label{eq9}
S=k\int d^{4}x\sqrt{g}\Theta.
\end{equation}
We can see clearly that this functional obeys the requirement of
being invariant under general coordinate transformations because the
volume element is the Riemannian one and the quantity $\Theta $ is a
scalar, in this case, the scalar of volume expansion. The meaning of
that functional is the following: it has the same mathematical form
as the measure of the linear deformation of a "fluid". Now, the
scalar of expansion is given by $\Theta =-\lambda R_{ab}\xi
^{a}\xi^{b}$ in the conditions already cited (for a congruence of
timelike curves, i. e., if the fundamental particles of the lattice
are massive). The spacetime metric can be decomposed as
$g_{ab}=h_{ab}+\xi_{a}\xi_{b}$. Then, the scalar of expansion can be
rewritten as $\Theta\sim -\tau R_{ab}g^{ab}\equiv -\tau R$.
Substituting this result in the functional defined above we arrive
at
\begin{equation}\label{eq10}
S\sim -k\int d^{4}x\tau \sqrt{g}R,
\end{equation}
which is just the Einstein-Hilbert action multiplied by the factor
$\tau $ (parameter of the curves). Nevertheless, in the standard
minimization procedure we make a variation of the action with the
fields to get the equations of motion. The parameter $\tau $ does
not have a functional variation and, therefore, the equations of
motion remain unchanged. The conclusion of this section is that when
we minimize the Einstein-Hilbert action we are indeed looking for
deformations of a "fluid" that minimize the functional of
deformation described above.

\section{Third sign: The Newton's law and the Hooke's law}

Regarding the conclusions of the sections above it is natural to ask
if the spacetime obeys some mathematical relation similar to the
equations describing phenomena related with material bodies. The
answer is positive as we will see. Consider that the spacetime is a
special kind of deformable body. We will postulate that for linear
deformations (for small dislocations of the "constituents" of that
body) we can write a Hooke's law that links the deformation tensor
$\varepsilon ^{ab}$ to the tensions $\tau ^{ab}$ applied on the body
in discussion (the spacetime). The tensors $\varepsilon ^{ab}$ and
$\tau ^{ab}$ are symmetric, in analogy with the definitions of these
objects in mechanics of the deformable bodies. Then, the Hooke's law
is:
\begin{equation}\label{eq11}
\tau
^{ab}=-k\varepsilon ^{ab}.
\end{equation}
The deformation tensor $\varepsilon ^{ab}$  is defined, in this
case, as the symmetrical part in the kinematical decomposition of
the tensor $B^{ab}=\nabla ^{a}\kappa^{b}$ in the construction of the
Landau-Raychaudhuri equation. In this way, its trace will obeys the
following rule:
\begin{equation}\label{eq12}
Tr\varepsilon ^{ab}=\varepsilon _{a}^{a}\equiv \Theta =-\lambda
R_{ab}\kappa
^{a}\kappa ^{b}.
\end{equation}
Note the type of relation between the deformation tensor
$\varepsilon ^{ab}$ and the Ricci tensor $R^{ab}$: it is the trace
of the deformation tensor that is linked to the Ricci tensor. This,
in a sense, denotes that the deformation tensor is an object
"bigger than" the Ricci tensor. Taking the trace in the expression
for the Hooke's law we obtain
\begin{equation}\label{eq13}
\tau _{a}^{a}=k\lambda R_{ab}\kappa ^{a}\kappa
^{b},
\end{equation}
where we used the relation for the trace of the deformation tensor.
We see now that the side of the deformations in the Hooke's law is
related with the geometry of the spacetime. The side of the tensions
must be, therefore, related to the material content that produces
the tensions on the spacetime. This is a reasonable assumption in
the sense that if we want to equate in the same way the two sides of
the Hooke's law we must require that the trace of the tension tensor
satisfies $\tau _{a}^{a}=\lambda T_{ab}\kappa ^{a}\kappa ^{b}$. This
is not difficult to accept if we remember the relation between the
deformation tensor and the Ricci tensor discussed above. Another way
to see this result is by the substitution of $\kappa ^{a}\kappa
^{b}$ by the metric of the spacetime together the quantities that
define projections in this spacetime. This will result in a trace
as we want. The parameter $\lambda $ enters in the expression by
dimensional reasons. Because of this we see that the tensor of
tension is, in the same way as the deformation tensor, an object
"bigger than" $T^{ab}$. Equating in this way deformations and
tensions we arrive at
\begin{equation}\label{eq14}
\left( R_{ab}-k^{-1}T_{ab}\right) \kappa ^{a}\kappa
^{b}=0,
\end{equation}
that is valid for all null vector $\kappa^{a}$, i. e.,
\begin{equation}\label{eq15}
R_{ab}-k^{-1}T_{ab}=f(g)g_{ab}.
\end{equation}
The function $f(g)$ again depends only on the metric $g_{ab}$ of the
spacetime and can be determined by requiring the disappearance of
the covariant divergence of the geometric part. But this only
happens if we require in addition the validity of $\nabla ^{a}T_{ab}=0$, i.
e., the quantity $T_{ab}$ must satisfies an equation of conservation.
Note that we have assumed a null-like congruence in discussing this
signal. Concluding, we obtain the equation
\begin{equation}\label{eq16}
R_{ab}-\frac{1}{2}Rg_{ab}\pm
\Lambda g_{ab}=k^{-1}T_{ab},
\end{equation}
which is just the Einstein's equation if we identify $T_{ab}$ with
the energy-momentum tensor. The identification is correct because
$T_{ab}$ is a symmetric object due to the symmetry of $R_{ab}$ and
it obeys a conservation law. If we take seriously these analogies we
are forced to declare that the Hooke's law for the spacetime is, in
this viewpoint, more fundamental than the Einstein's equation. In
other words, \textit{the Newton's law of gravitation comes from the Hooke's
law describing the deformations of the spacetime}.

\section{Conclusions and Perspectives}

In this work we discussed the relations between the idea of
spacetime curvature and deformations of solids. Interesting signs
can be constructed in analogy with the physics of material bodies.
The first sign shows that if we deform the spacetime volume using
Lie propagation through integral curves we obtain the geometrical
part of the Einstein's equation. In the second sign, we proof a
relation between the Einstein-Hilbert action and the volume
deformation of a kind of "solid". In the third sign, we postulate
the validity of a Hooke's law for the spacetime and show that the
Newton' law of gravitation appears through Einstein's equation. In
all of these signs, the Landau-Raychaudhuri equation plays important
role in the description of the spacetime as fulfilled by a
congruence of curves. Nevertheless, we make use of a time-like
congruence in just one sign while in the others we use null-like
congruences. This means that it is important to decide if the
fundamental particles of the lattice are massive or non-massive in
order to establish the characteristics of the proposed lattice.
Until now, there is no fundamental reason to choose one or other
sort of congruence. We regard all of these signs as important steps
in order to compose the idea of this paper. This question will be
better addressed in a forthcoming paper.

The authors would like to thank Funda\c{c}\~{a}o Cearense de apoio
ao Desenvolvimento Cient\'{\i}fico e Tecnol\'{o}gico (FUNCAP) and
Conselho Nacional de Desenvolvimento Cient\'{\i}fico e
Tecnol\'{o}gico (CNPq) for financial support.


\begin{thebibliography}{99}
\addcontentsline{toc}{chapter}{Bibliografia}

\bibitem{sakharov} A. D. Sakharov, "Vacuum quantum fluctuations in curved space and the theory of gravitation", Soviet Physics Doklady, \textbf{12} (1968) 1040.
\bibitem{gravity_estocastica} B.L. Hu, E. Verdaguer, "Stochastic gravity: Theory and applications", Living Rev. Relativity, \textbf{7} (2004) 3.
\bibitem{jacobson} T. Jacobson, "Thermodynamics of spacetime: The Einstein equation of state", Phys. Rev. Lett. \textbf{75} (1995) 1260.
\bibitem{lan} L. Landau and E. M. Lifshitz, Classical theory of fields, Pergamom Press, Oxford, UK, 1975.
\bibitem{ray} A. Raychaudhuri, Phys. Rev. \textbf{98} (1955) 1123.
\bibitem{wald} R. M. Wald, General Relativity, Chicago, USA: Univ. Press (1984)



\end{thebibliography}
\end{document}